\documentclass[aoas,MSNbibl,nameyear,rotating,dvips]{arximspdf}
\usepackage[ruled]{algorithm2e}
\usepackage{dcolumn}
\usepackage{graphicx}

\doi{10.1214/14-AOAS716} 
\volume{8}
\issue{2}
\pubyear{2014}
\firstpage{1119}
\lastpage{1144}

\makeatletter
\let\my@algocf@latexcaption\algocf@latexcaption
\let\my@addcontentsline\addcontentsline
\long\def\algocf@latexcaption#1[#2]#3{%
\def\addcontentsline##1##2##3{}%
\my@algocf@latexcaption{#1}[#2]{#3}%
\global\let\addcontentsline\my@addcontentsline%
}
\newcolumntype{d}[1]{D{.}{.}{#1}}
\newcommand{\rrvert}{\vert}
\newcommand{\llvert}{\vert}
\newcommand{\PP}{\mathrm{P}}
\newcommand{\VV}{\mathrm{V}}
\makeatother

\begin{document}
\begin{frontmatter}

\title{A statistical approach to the inverse problem in~magnetoencephalography}
\runtitle{MEG inverse problem}

\begin{aug}
\author[A]{\fnms{Zhigang} \snm{Yao}\corref{}\thanksref{t1,m1}\ead[label=e1]{zhigang.yao@epfl.ch}}
\and
\author[B]{\fnms{William F.} \snm{Eddy}\thanksref{m2}\ead[label=e2]{bill@stat.cmu.edu}}
\runauthor{Z. Yao and W. F. Eddy}
\affiliation{Ecole Polytechnique F\'{e}d\'{e}rale de Lausanne\thanksmark{m1} and Carnegie Mellon University\thanksmark{m2}}
\address[A]{Section de Math\'{e}matiques\\
Ecole Polytechnique F\'{e}d\'{e}rale de Lausanne\\
EPFL Station 8, 1015 Lausanne\\
Switzerland\\
\printead{e1}} 
\address[B]{Department of Statistics\\
Carnegie Mellon University\\
Pittsburgh, Pennsylvania 15213\\
USA\\
\printead{e2}}
\end{aug}
\thankstext{t1}{Supported in part by NSF SES-1061387 and NIH/NIDA R90 DA023420.}

\received{\smonth{1} \syear{2012}}
\revised{\smonth{1} \syear{2014}}

%
\begin{abstract}
Magnetoencephalography (MEG) is an imaging technique used to measure
the magnetic field outside the human head produced by the electrical
activity inside the brain. The MEG inverse problem, identifying the
location of the electrical sources from the magnetic signal
measurements, is ill-posed, that is, there are an infinite number of
mathematically correct solutions. Common source localization methods
assume the source does not vary with time and do not provide estimates
of the variability of the fitted model. Here, we reformulate the MEG
inverse problem by considering time-varying locations for the sources
and their electrical moments and we model their time evolution using a
state space model. Based on our predictive model, we investigate the
inverse problem by finding the posterior source distribution given the
multiple channels of observations at each time rather than fitting
fixed source parameters. Our new model is more realistic than common
models and allows us to estimate the variation of the strength,
orientation and position. We propose two new Monte Carlo methods based
on sequential importance sampling. Unlike the usual MCMC sampling
scheme, our new methods work in this situation without needing to tune
a high-dimensional transition kernel which has a very high cost. The
dimensionality of the unknown parameters is extremely large and the
size of the data is even larger. We use Parallel Virtual Machine (PVM)
to speed up the computation.
\end{abstract}

%

\end{frontmatter}

\section{Introduction}\label{sec1}
\subsection{The basics of magnetoencephalography (MEG)}\label{sec1.1} The anatomy of
the brain has been studied intensively
for millennia, yet how the brain functions is still not well
understood. The neurons in the brain produce macroscopic electrical
currents when the brain functions,\vadjust{\goodbreak} and those synchronized neuronal
currents in the gray matter of the brain induce extremely weak
magnetic fields (10--100~femto-Tesla) outside the head. The
comparatively recent development of the Superconducting Quantum
Interference Device (SQUID) makes it possible to detect those magnetic
signals. MEG is an imaging technique using SQUIDs to measure the
magnetic signals outside of the head produced by the electrical
activity inside the brain [\citet{Cohen1968}]. The primary sources
are electric currents within the dendrites of the large pyramidal cells
of activated neurons in the human cortex, generally formulated as a
mathematical point current dipole. Such focal brain activation can be
observed in epilepsy, or it can be induced by a stimulus in
neurophysiological or neuropsychological experiments.
Due to its
noninvasiveness (it is a~completely passive measurement method) and
its impressive temporal resolution (better than 1 millisecond,
compared to 1 second for functional magnetic resonance imaging, or to
1 minute for positron emission tomography), and due to the fact that
the signal it measures is a direct consequence of neural activity, MEG
is a near optimal tool for studying brain activity,
such as assisting surgeons
in localizing a pathology, assisting researchers in determining brain function,
neuro-feedback and others.
The skull and the tissue surrounding the brain affect the magnetic
fields measured by MEG much less than the electrical impulses measured
by electroencephalography (EEG). This means that MEG has higher
localization accuracy than the EEG and it allows for a more reliable
localization of brain function [\citet{Hamalainen1993}, \citet
{okada1999}]. MEG has recently been used in the evaluation of epilepsy,
where it reveals the exact location of the abnormalities, which may
then allow physicians to find the cause of the seizures [\citet
{Barkley2003}]. MEG is also reference free, so that the localization of
sources with a given precision is easier for MEG than it is for EEG
[\citet{Kristeva1997}].
The computation associated with
estimating the electric source from the magnetic measurement is a
challenging problem that needs to be solved to allow high temporal and
spatial resolution imaging of the dynamic activity of the human brain.


\subsection{Forward and inverse MEG problem}\label{sec1.2}
The MEG signals derive from the \textit{primary} current (the net
effect of ionic currents flowing in the dendrites of neurons) and the
\textit{volume} current (i.e., the additive ohmic current set up in
the surrounding medium to complete the electrical circuit). If the
electrical source is known and the head model [\citet{Kybic2006}]
is specified (e.g., a sphere with homogeneous conductivity), then the
``forward problem'' is to compute the electric field $\mathbf{E}$ and
the magnetic field $\mathbf{B}$ from the source current $\mathbf{J}$.
The calculation uses Maxwell's equations [see, e.g., \citet{Griffiths1999}],
\begin{eqnarray*}
\nabla\cdot\mathbf{E}&=&\rho/\varepsilon_{0},
\\
\nabla\times\mathbf{E}&=&-\partial\mathbf{B}/\partial t,
\\
\nabla\cdot\mathbf{B}&=&0,
\\
\nabla\times\mathbf{B}&=&\mu_{0}(\mathbf{J}+\varepsilon_{0}
\,\partial\mathbf{E}/\partial t),
\end{eqnarray*}
where $\varepsilon_{0}$ and $\mu_{0}$ are the permittivity and
permeability of a vacuum, respectively, and $\rho$ is the charge
density. The total current $\mathbf{J}$ consists of the primary current
$\mathbf{J}^{\PP}$ plus the volume current $\mathbf{J}^{\VV}$. The source
activity in the brain corresponds to the primary current. Under
reasonable assumptions [see \citet{Hamalainen1993}], the volume
current $\mathbf{J}^{\VV}$ is not included in the analysis because of its
diffuse nature. The terms $\partial\mathbf{B}/\partial t$ and
$\partial\mathbf{E}/\partial t$ in Maxwell's equations can be ignored
by assuming that the magnetic field varies relatively slowly in time.
Rather than working with continuous electric current, the most
frequently used computational model assumes that the electric current
can be thought of as an electric dipole; this model is called an
equivalent current dipole (ECD); see, for example, \citet{Hamalainen1993}. From the perspective of an ECD, a dipole has
location, orientation and magnitude; the magnetic field generated by
this dipole can explain the MEG measurements. In addition, there is a
version of an ECD model assuming multiple dipoles; from Maxwell's
equations it is easy to see that this model is simply the sum of the
models for each ECD. Such an ECD models a large number of dipoles
located at fixed places over the cortical surface. In neuroscience, it
is believed that typical MEG data should be explained by only a few
dipoles (less than 10), and different criteria or algorithms are used
to minimize the number of dipoles in various ECD models; we discuss
some of these models in Section~\ref{sec1.3}.
We assume that $\mathbf{E}$ is generated by~$\mathbf{J}^\PP$, which in
turn comes from the sum of $N$ localized current dipoles at locations~$\mathbf{r}_{n}$,
\[
\mathbf{J}_{n}^{\PP}(\mathbf{r})=Q_{n}\delta(
\mathbf{r}-\mathbf{r}_{n}),\qquad n=1,\ldots,N,
\]
where $\delta(\cdot)$ is the Dirac delta function. The $Q_{n}$ is a
charged dipole at the point $\mathbf{r}_{n}$ in the brain volume $\Omega
$. Using the quasi-static approximation to Maxwell's equations (i.e.,
ignoring the partial derivatives with respect to time) given in
\citet{Sarvas1987}, the magnetic field $\mathbf{B}$ at location
$\mathbf{r}$ of a current dipole at $\mathbf{r}_{n}$ can be calculated
by the Biot--Savart equation,
\[
\mathbf{B}_n(\mathbf{r})=\frac{\mu_{0}}{4\pi}\int_{\Omega}
\frac{\mathbf{J}^{\PP}(\mathbf{r}_{n})\times(\mathbf{r}-\mathbf
{r}_{n})}{\llvert\mathbf{r}-\mathbf{r}_{n}\rrvert^{3}}\,d\mathbf{r}_{n}.
\]
%
In the case of multiple current dipoles, the induced magnetic fields
simply add up.

The ``inverse problem'' comes from the forward model; we want to
estimate the dipole parameters from the observed magnetic signal. The
difficulty is that there is not a unique solution; there are infinitely
many different sources within the skull that produce the same observed
data. The goal is to find a meaningful solution among the many
mathematically correct solutions. There are three key steps to any
source localization algorithm in MEG. First, define the solution space
and the parameter space of the electric source. Second, calculate the
magnetic field given the information about the head model. Third,
according to what criterion the solution must satisfy, perform an
extensive search for the solution. Methods of finding the source from
the observed MEG signal have been extensively exploited during the past
two decades, mostly centered on finding a single estimate of the
source. Some of these methods are briefly described in the next
subsection. However, finding the distribution of the source in space
and (particularly) in time is still a problem requiring investigation.

\subsection{Existing source localization methods}\label{sec1.3}
Most methods for localizing electrical sources in MEG
assume that the electrical sources in the brain do not include a
temporal component. The data are used to estimate the source
parameters at each time point; there is no relation to the
estimates for the previous time. This is not the same as assuming the
quasi-static approximation to Maxwell's equations. Therefore, those
existing methods are restricted to fixed dipole assumptions and are
also not able to provide estimates of the variability of source
activity. The minimum norm estimate (MNE)
[\citet{Hamalainen1994}] is a regularization method based on the
$L_{2}$ norm. The $L_{1}$ norm regularization yields the minimum
current estimates (MCE) [\citet{Uutela1999}]. The LORETA approach
[\citet{Mattouta2006}] is a special case of weighted MNE. The Multiple
Signal Classification method (MUSIC) [\citet{Mosher1992}] searches for
a single-dipole model through a three-dimensional head volume and
computes projections onto an estimated signal subspace.
The source locations are then found as the 3-D locations where the
source model
gives the best projections onto the subspace.
The beamformer methods [\citet{VanVeen1992}]
ignore the ill-posed inverse problem and instead only estimate the
current at several fixed locations. Bayesian approaches to the MEG
inverse problem try to find the posterior distribution of the dipole
parameters [\citet{Bertrand2001}, \citet{Schmidt1999}].

The methods mentioned briefly above (MNE, MCE, etc.) have been widely
used and produce apparently meaningful solutions; however they have
overly restrictive assumptions and lack estimates of the variability of
source estimates. By assuming a static localized dipole, these methods
are limited in their ability to incorporate problem-specific anatomical
or physiological information. It is quite reasonable to consider that
the source is time-varying rather than fixed, in which case the noise
reduction obtained by averaging over consecutive observations in time
is problematic. By utilizing a time-varying source model, we will be
able to investigate the distribution of the source at each time point
and provide estimates of the variability. Following this idea, the time
evolution of the source is modeled by a state space model. Our goal is
to find the posterior distribution of the source parameters. Our
reformulation of the inverse problem leads to a predictive model of the
dipole. It turns out that the posterior source distribution from our
predictive model can be interpreted as a statistical solution to the
MEG inverse problem.

\subsection{Outline of this paper}\label{sec1.4}
In Section~\ref{sec2} we develop a time-varying source model for the MEG inverse problem.
Rather than attempting to ``solve'' the inverse problem we try to
develop estimates of the dipole parameters using a spatio-temporal
model. In Section~\ref{sec3} the difficulty of using Markov chain Monte Carlo
(MCMC) methods for generating samples from the time-varying model is
explained. Then, we introduce the standard Sequential Importance
Sampling (SIS) technique. Next, two further Monte Carlo methods are
described: (1) the regular SIS method with rejection, and (2) the
improved SIS method with resampling. A~simulation study is described in
Section~\ref{sec4}. We describe our use of the Parallel Virtual Machine (PVM)
software to speed up the computations in Section~\ref{sec4.2}. We believe this
is the first application of parallel computational methods to the
problem. In Section~\ref{sec5} a real data application is presented. Section~\ref{sec6}
contains a short discussion and conclusion. 
\section{A probablistic rime-varying source model}\label{sec2}

Assume that the magnetic field data is measured from the $k$th sensor
at time $t$ as
\[
Y_{k,t}=\mathbf{B}_{k} \bigl(\mathbf{J}^{\PP}_{t}
\bigr)+\mathbf{U}_{k,t}, \qquad1\leq t\leq T, 1\leq k\leq L,
\]
where $\mathbf{U}_{k,t}\sim N(0, \sigma_{1}^{2})$ denotes the
observation noise that is assumed, for simplicity, to be Gaussian,
additive and homogeneous for all the sensors, and uncorrelated between
every pair of sensors. The assumption of normality is preferred due to
the fact that Gaussian sensor noise is present at the MEG sensors
themselves, and sensor noise is typically substantially smaller than
signals from spontaneous brain activity [\citet{Hamalainen1993}].
Although correlated sensor noise is more realistic than ``homogeneous''
sensors, it complicates the problem. Background noise and biological
noise can also drown out the brain activity of interest, but these are
all very difficult to incorporate. Besides some variation coming from
solving the inverse problem, most of the variation of the source
localization in MEG is due to the propagation of errors through
Maxwell's equations when solving the forward problem. In order to
control variation, we only work with a simple sensor structure.
Therefore, we write
\[
\mathbf{Y}_{t}=\mathbf{B} \bigl(\mathbf{J}^{\PP}_{t}
\bigr)+\mathbf{U}_{t}, \qquad1\leq t\leq T,
\]
where $\mathbf{Y}_{t}=(Y_{1,t},\ldots,Y_{L,t})^T$, $\mathbf{B}(\mathbf
{J}^{\PP}_{t})=(\mathbf{B}_{1}(\mathbf{J}^{\PP}_{t}),\ldots,\mathbf
{B}_{L}(\mathbf{J}^{\PP}_{t}))^T$ and $\mathbf{U}_{t}=(\mathbf
{U}_{1,t},\break  \ldots,\mathbf{U}_{L,t})^T$. Here, $\mathbf{U}_{t}\sim
\mathcal{N}(\mathbf{0},\bolds{\Sigma}_{1})$, where $\bolds{\Sigma
}_{1}=\operatorname{diag}[\sigma_{1}^{2},\ldots,\sigma_{1}^{2}]$.

The $\mathbf{B}_{k}(\mathbf{J}_{t}^{\PP})$, a function of the dipole with
parameter vector $\mathbf{J}_{t}^{\PP}$, is the physical approximation of
the Biot--Savart law in Section~\ref{sec1.2}.\vadjust{\goodbreak} We consider a current dipole
located within a horizontally layered conductor [\citet
{Hamalainen1993}]. The noiseless magnetic field at the $k$th sensor,
$\mathbf{B}_{k}$, is computed from the source $\mathbf
{J}_{t}^{\PP}=(\mathbf{p}_{t},\mathbf{q}_{t})$ at time $t$. The vector
$\mathbf{p}_{t}=({p_{1}}_{t},{p_{2}}_{t},{p_{3}}_{t})$ contains the
location parameters of the source and the vector $\mathbf
{q}_{t}=({q_{1}}_{t},{q_{2}}_{t},{q_{3}}_{t})$ contains the moments and
strength. Thus,
\begin{equation}
\label{biot} \mathbf{B}_{k} \bigl(\mathbf{J}^{\PP}_{t}
\bigr)=\frac{\mu_{0}}{4\pi}\frac{\mathbf{q}_{t}\times(\mathbf
{r}_{k}-\mathbf{p}_{t})\cdot\mathbf{e} }{\llvert\mathbf{r}_{k}-\mathbf
{p}_{t}\rrvert^{3}}.
\end{equation}
Here, $\mathbf{r}_{k}$ is the location of the $k$th sensor. Because the
magnetometers measure only the $z$ direction of the magnetic field
$\mathbf{B}$, $\mathbf{e}=(0,0,1)$, a unit vector, is used to find the
$z$ component of $\mathbf{B}$. Conventionally, $z$ is perpendicular to
the surface of the skull.

To specify the \textit{prior}, the time evolution of the current dipole
$\mathbf{J}^{\PP}_{t}$ is modeled by a state space model. A number of other authors have also proposed stating the MEG inverse problem as a Bayesian dynamic
model; see \citet{SomVouKai03}; \citeauthor{Cametal08} (\citeyear{Cametal08,Cametal});
\citeauthor{Soretal09} (\citeyear{Soretal09,Soretal13});
\citet{Miaetal13}.
We could choose
any of a large variety of state space models but, for theoretical and
computational simplicity, we have chosen a six-dimensional first-order
autoregression:
\[
\mathbf{J}^{\PP}_{t}=\mathbf{m}_{\mathrm{com}}+\bolds{\rho}
\bigl(\mathbf{J}^{\PP}_{t-1}-\mathbf{m}_{\mathrm{com}} \bigr)+
\mathbf{V}_{t}, \qquad1\leq t\leq T,
\]
where $\mathbf{V}_{t}\sim\mathcal{N}(\mathbf{0},\bolds{\Sigma}_{2})$
denotes the state evolution noise. We note that three of the parameters
give the spatial location, so this is implicitly a space--time model.
Previous work using a space--time model [\citet{Ou2009}] used a
novel mixed $L_1L_2$-norm estimate for the dipole parameters based on a
linear regression model.
\citet{Junetal05} have also used MCMC methods for sampling from the posterior of a spatiotemporal
Bayesian dynamic model.
To reduce the number of parameters, and hence
the amount of variation in our estimates, we assume the dipole
parameters are uncorrelated. That is, we assume that $\bolds{\Sigma
}_{2}=\operatorname{diag}[\sigma_{11}^{2},\sigma_{22}^{2},\ldots,\sigma
_{66}^{2}]$ is a known $6$ by $6$ diagonal matrix and $\sigma_{ii}^{2}$
is the variance of the $i$th source parameter. The\vspace*{2pt} parameter vector
$\mathbf{m}_{\mathrm{com}}$ is a constant associated with the source $\mathbf
{J}^{\PP}_{t}$ for $1\leq t\leq T$. The initial state is chosen as
$\mathbf{J}^{\PP}_{0} \sim\mathcal{N}(\mathbf{m}_{\mathrm{ini}},\bolds{\Sigma
}_{2})$, where $\mathbf{m}_{\mathrm{ini}}$ is also a constant parameter vector.
Both $\mathbf{m}_{\mathrm{ini}}$ and $\mathbf{m}_{\mathrm{com}}$ are specified in
advance. The known diagonal matrix $\bolds{\rho}=\operatorname{diag}[\rho
_{1},\rho_{2},\ldots,\rho_{6}]$ is $6$ by $6$. Its main diagonal
represents the autoregressive coefficients. Hence, at any time~$t$,
$\mathbf{J}^{\PP}_{t}$~or~$(\mathbf{p}_{t},\mathbf{q}_{t})$ contains\vspace*{1pt} the
parameters of interest and $\mathbf{Y}_{t}=(Y_{1,t},\ldots,Y_{L,t})$ is
the (very noisy) data collected from all sensors. Both $\{\mathbf
{J}^{\PP}_{t}\}_{t=0}^{T}$ and $\{Y_{k,t}\}_{t=1}^{T}$ are assumed to
have the following Markov properties:
\begin{longlist}[(iii)]
\item[(i)] The $\mathbf{J}^{\PP}$ is a first order Markov process. The
distribution of each state $\mathbf{J}^{\PP}_{t}$ only depends on its own
previous state $\mathbf{J}^{\PP}_{t-1}$,
\begin{equation}
\label{eq_i} p \bigl(\mathbf{J}^{\PP}_{t}|
\mathbf{J}^{\PP}_{0}, \mathbf{J}^{\PP}_{1},\ldots,\mathbf{J}^{\PP}_{t-1} \bigr)=p \bigl(\mathbf{J}^{\PP}_{t}|
\mathbf{J}^{\PP}_{t-1} \bigr)\vadjust{\goodbreak}
\end{equation}
(we are using $p$ as a generic symbol for a probability distribution;
the two $p$'s in this equation are not the same function).

\item[(ii)] The process $Y_{k,t}$ (for any $1\leq k\leq L$) is also a
Markov process with respect to the history of $\mathbf{J}^{\PP}$. The
density of $Y_{k,t}$ conditioned on $\{\mathbf{J}^{\PP}_{t}\}_{0}^{t}$ satisfies
\[
\label{eq_ii} f \bigl(Y_{k,t}|\mathbf{J}^{\PP}_{0},
\mathbf{J}^{\PP}_{1},\ldots,\mathbf{J}^{\PP}_{t}
\bigr)=f \bigl(Y_{k,t}|\mathbf{J}^{\PP}_{t} \bigr)
\]
(again $f$ is a generic symbol, in this case, for a joint density
function).

\item[(iii)] When conditioned on its own history, the unknown $\mathbf
{J}^{\PP}_{t}$ does not depend on past measurements. The distribution of
$\mathbf{J}^{\PP}_{t}$ based on $\mathbf{Y}^{k}=(Y_{k,1},\ldots,\break Y_{k,t-1})$ and $\mathbf{J}^{\PP}_{t-1}$ is
\begin{equation}
\label{eq_iii} g \bigl(\mathbf{J}^{\PP}_{t}|
\mathbf{J}^{\PP}_{t-1}, \mathbf{Y}^{k} \bigr)=p \bigl(
\mathbf{J}^{\PP}_{t}| \mathbf{J}^{\PP}_{t-1}
\bigr),\qquad t>0
\end{equation}
[the right-hand side of equation (\ref{eq_iii}) in (iii) is the same as
the right-hand side of equation (\ref{eq_i}) in (i)].
The transition kernel, $p(\mathbf{J}^{\PP}_{t}|\mathbf{J}^{\PP}_{t-1})$, is
defined here as a first order Markov process in the state space model
above. For a more complex model it could be a higher order Markov
process. The choice of more realistic models for this process [e.g., in
the situation where the magnetic signal is a response to a stimulus,
the source variance might change much more rapidly immediately after
the stimulus than before it; the joint density $f(Y_{k,t}|\mathbf
{J}^{\PP}_{t})$ for any $1\leq k\leq L$ may also vary in time since not
all the measurements can be carried out simultaneously] is not the aim
of this paper. 

Of interest at any time $t$ is the posterior distribution of $\mathcal
{J}^{\PP}_{t}=(\mathbf{J}^{\PP}_{0},\ldots,\mathbf{J}^{\PP}_{t})$. Let
$\mathcal{Y}_{\mathrm{obs}}^{t}=(\mathbf{Y}_{1},\ldots,\mathbf
{Y}_{L})=({Y_{1,1},\ldots,Y_{1,t}},\dots, {Y_{L,1},\ldots,Y_{L,t}})$ be
the magnetic measurements, accordingly. By taking all the previous
\textit{prior} information and the three assumptions [(i), (ii), (iii)]
above into account, our problem can be stated as finding the target
distribution, $p(\mathcal{J}^{\PP}_{t}|\mathcal{Y}_{\mathrm{obs}}^{t})$, given
$\mathcal{Y}_{\mathrm{obs}}^{t}$. By Bayes' theorem, we have
\begin{eqnarray}
\label{likelihood} p \bigl(\mathcal{J}^{\PP}_{t}|
\mathcal{Y}_{\mathrm{obs}}^{t} \bigr)&\propto& f \bigl(
\mathcal{Y}_{\mathrm{obs}}^{t}|\mathcal{J}^{\PP}_{t}
\bigr)p \bigl(\mathcal{J}^{\PP}_{t} \bigr)
\nonumber\\[-8pt]\\[-8pt]
&=& \Biggl[\prod_{s=1}^{t}\prod
_{k=1}^{L}f \bigl(Y_{k,s}|
\mathbf{J}^{\PP}_{s} \bigr) \Biggr] \Biggl[\prod
_{s=1}^{t}p \bigl(\mathbf{J}^{\PP}_{s}|
\mathbf{J}^{\PP}_{s-1} \bigr) \Biggr]p \bigl(
\mathbf{J}^{\PP}_{0} \bigr).\nonumber
\end{eqnarray}
This framework is based on a one-source model ($N=1$). It can be easily
extended to a multiple-source model because the fields generated by
distinct sources simply add up. Because it is a high-dimensional
distribution ($1\leq t\leq T$, $T$ is very large) and inherently
complicated, sampling from the posterior is difficult. We have chosen
to use MCMC methods but they are also complex and are very hard to
implement. As we will show later, obtaining $p(\mathcal
{J}^{\PP}_{t}|\mathcal{Y}_{\mathrm{obs}}^{t})$ can be achieved dynamically by
computing the $p(\mathbf{J}^{\PP}_{u}|\mathcal{Y}_{\mathrm{obs}}^{u})$ at each
time point $1 \leq u \leq t$. These calculations have to be repeated
for each $t\leq T$.
\end{longlist}

\section{Solving the MEG inverse problem}\label{sec3}
\subsection{The difficulty of solving the time-varying model}\label{sec3.1}
A problem with MCMC methods (e.g., Metropolis--Hastings) for getting
joint posterior samples from $p(\mathcal{J}^{\PP}_{t}|\mathcal
{Y}_{\mathrm{obs}}^{t})$ occurs when there are a large number of states because
it is difficult to find a joint transition kernel which could be used
in an MCMC sampler. However, the goal of getting $p(\mathcal
{J}^{\PP}_{t}|\mathcal{Y}_{\mathrm{obs}}^{t})$ can be achieved by sampling from
the distribution $p(\mathbf{J}^{\PP}_{s}|\mathcal{Y}_{\mathrm{obs}}^{s})$ for each
state $s$ ($1\leq s\leq t$) separately and the entire outcome could be
regarded as the sample from the joint distribution. Gibbs sampling can
be used for this restricted goal, but because of
the nonlinearity of the model [equation (\ref{biot})], it is not easy
to sample from $p(\mathbf{J}^{\PP}_{t}|\mathbf{J}^{\PP}_{s \neq t},\mathcal
{Y}_{\mathrm{obs}}^{t})$. One way to alleviate the difficulty is to insert some
kind of Metropolis sampler into a Gibbs sampling scheme for each
conditional distribution. When we insert a random-walk Metropolis
algorithm, where the move depends only on its own state, into the Gibbs
sampler, we call it a \textit{random-walk MCMC within Gibbs} sampler, and
when we insert a hybrid Metropolis algorithm, where the move may depend
on other states, into the Gibbs sampler, we call it a \textit{hybrid MCMC
within Gibbs} sampler.

The key to \textit{random-walk MCMC within Gibbs} is to propose a
candidate $\mathbf{J}^{\PP^{*}}_{t}\sim\mathcal{N}(\mathbf
{J}^{\PP}_{t},\bolds{\Sigma}_{3})$ for each $t$ ($1 \leq t \leq T$),
where $\bolds{\Sigma}_{3}=\operatorname{diag}[\tau_{1}^{2},\tau_{2}^{2},\ldots,\tau_{6}^{2}]$ is a $6$ by $6$ diagonal matrix, and accept $\mathbf
{J}^{\PP^{*}}_{t}$ if the acceptance ratio
\[
\alpha_{t}=\frac{\prod_{k=1}^{L}f(Y_{k,t}|\mathbf
{J}^{\PP^{*}}_{t})p(\mathbf{J}^{\PP^{*}}_{t}|\mathbf{J}^{\PP}_{t-1})p(\mathbf
{J}^{\PP}_{t+1}|\mathbf{J}^{\PP^{*}}_{t})}{\prod_{k=1}^{L}f(Y_{k,t}|\mathbf
{J}^{\PP}_{t})p(\mathbf{J}^{\PP}_{t}|\mathbf{J}^{\PP}_{t-1})p(\mathbf
{J}^{\PP}_{t+1}|\mathbf{J}^{\PP}_{t})} \geq\mathcal{U}(0,1),
\]
where $\mathcal{U}(0,1)$ is the uniform distribution. The problem is
that $\mathcal{N}(\mathbf{J}^{\PP}_{t},\bolds{\Sigma}_{3})$ is not a
good proposal for $\mathbf{J}^{\PP^{*}}_{t}$ (i.e., we almost always
reject the proposal) and this can not be solved by extensively tuning
$\bolds{\Sigma}_{3}=\operatorname{diag}[\tau_{1}^{2},\tau_{2}^{2},\ldots,\tau
_{6}^{2}]$ in most practical cases if the dimension of the states is
very high. A local linear approximation might be considered, such as
performing a Taylor expansion on the joint density function
$f({Y}_{k,t}|\mathbf{J}^{\PP}_{t})$ and truncating high order terms. The
resultant can then be incorporated into the proposal distribution.
However, such linearization is not easy due to the highly complex
function $f({Y}_{k,t}|\mathbf{J}^{\PP}_{t})$; moreover, the extra work of
a Taylor expansion might be unnecessary if we only need an efficient
sampling scheme in high dimensions.

The \textit{hybrid MCMC within Gibbs} improves upon the random-walk
MCMC within Gibbs when the target distribution is not able to be
captured by a simple random walk. In \citet{Gelman1996}, a full
conditional prior (hybrid MCMC) was proposed. Similar work can also be
found in \citet{Carter1994}, where a single move blocking strategy
was developed but bad convergence behavior was discovered. \citet
{Gamerman1998} suggested a reparameterization of the model to a prior
independent system of disturbances and built a proposal by a weighted
least squares algorithm, however, the reparameterization resulted in
quadratic computational time. \citet{Leonhard1999} suggested an
autoregressive prior where the ``conditional prior'' is drawn
independently of the current state but, in general, depends on other
states. Here, our hybrid MCMC within Gibbs is built on a single move
proposal, that is, $\mathbf{J}^{\PP^{*}}_{t}$ is\vspace*{1pt} proposed from the
distribution of $p(\mathbf{J}^{\PP}_{t}|\mathbf{J}^{\PP}_{s\neq t})$ which
can be further reduced to $p(\mathbf{J}^{\PP}_{t}|\mathbf
{J}^{\PP}_{t-1},\mathbf{J}^{\PP}_{t+1})$ due to\vspace*{-1pt} the Markov property. One
way to update $\mathbf{J}^{\PP}_{t}$ is to use a proposal
\[
\mathbf{J}^{\PP^{*}}_{t} \sim\mathcal{N} \bigl(\bolds{\rho}
\bigl(\mathbf{J}^{\PP}_{t-1}-\mathbf{J}^{\PP}_{t+1}
\bigr)+ \bigl(\mathbf{I}-\bolds{\rho}\bolds{\rho}' \bigr) \bigl(
\mathbf{I}+\bolds{\rho}\bolds{\rho}' \bigr)^{-1}
\mathbf{m}_{\mathrm{com}},\bolds{\Sigma}_{2} \bigl(\mathbf{I}+\bolds{
\rho}\bolds{\rho}' \bigr)^{-1} \bigr).
\]
The acceptance ratio then reduces to
\[
\alpha_{t}=\frac{\prod_{k=1}^{L}f(Y_{k,t}|\mathbf{J}^{\PP^{*}}_{t})}{\prod
_{k=1}^{L}f(Y_{k,t}|\mathbf{J}^{\PP}_{t})}.
\]

The performance of a single move could be extended to a block move by
sampling a block of states at the same time based on other states.
Similarly, the $\mathbf{J}^{\PP^{*}}_{r},\ldots,\mathbf{J}^{\PP^{*}}_{s}$
come from the conditional proposal
\[
p \bigl(\mathbf{J}^{\PP}_{r},\ldots,\mathbf{J}^{\PP}_{s}|
\mathbf{J}^{\PP}_{1,\ldots,T}/ \bigl(\mathbf{J}^{\PP}_{r},\ldots,\mathbf{J}^{\PP}_{s} \bigr) \bigr),
\]
where $r<s$ and $\mathbf{J}^{\PP}_{1,\ldots,T}/(\mathbf{J}^{\PP}_{r}, \ldots,\mathbf{J}^{\PP}_{s})$ means a collection of $\mathbf{J}^{\PP}_{1},\ldots,\mathbf{J}^{\PP}_{r-1}, \mathbf{J}^{\PP}_{s+1},\break\ldots,\mathbf{J}^{\PP}_{T}$. Thus, the acceptance ratio becomes
\[
\alpha_{t}=\frac{\prod_{k=1}^{L}\prod_{t=r}^{s}f(Y_{k,t}|\mathbf
{J}^{\PP^{*}}_{t})}{\prod_{k=1}^{L}\prod_{t=r}^{s}f(Y_{k,t}|\mathbf{J}^{\PP}_{t})}.
\]
Although the block move provides a considerable improvement in the
situation where a single move has poor mixing behavior, \citet
{Carter1994} observed bad mixing and convergence behavior in the
blocking strategy.

Recently developed adaptive samplers [\citet{Andrieu2008},
\citet{RR2009}] might help find the transition kernel within a
Gibbs sampler, but these methods are computationally inefficient in
high dimension. In addition, although parallel tempering [\citet
{Srinivasan2002}] seems reasonable, finding the \mbox{temperature} is not
straightforward and significantly increases the computational cost.
Again, the MEG data set is extremely large; in particular, we collect
hundreds of channels of data at each time and we collect data for
hundreds of thousands of time points. It is quite
difficult to implement these methods since even a simple model has an
extremely large number of states. The computational burden is even more
substantial in the multiple-dipole case. 

\subsection{Sequential importance sampling (SIS)}\label{sec3.2}
Sequential importance sampling (SIS) [\citet{Liu1998}] is
advocated as a more practical tool for a dynamic system. As we\vspace*{1pt}
mentioned briefly in Section~\ref{sec2}, computing $p(\mathbf{J}^{\PP}_{u}|\mathcal
{Y}^{u}_{\mathrm{obs}})$ sequentially in $u$ for $1\leq u \leq t$ can lead to
$p(\mathcal{J}^{\PP}_{t}|\mathcal{Y}_{\mathrm{obs}}^{t})$. Consider\vspace*{1pt} $\pi_t(\mathbf
{J}^{\PP}_{t})=p(\mathbf{J}^{\PP}_{t}|\mathcal{Y}^{t}_{\mathrm{obs}})$; calculating
$p(\mathcal{J}^{\PP}_{t}|\mathcal{Y}_{\mathrm{obs}}^{t})$ or, equivalently,\vadjust{\goodbreak} $\pi
_t(\mathcal{J}^{\PP}_{t})$ can be achieved by performing the following
two processes in sequential order:
\begin{eqnarray}
\label{above} \pi_t \bigl(\mathbf{J}^{\PP}_{t}
\bigr)&=&\frac{f(\mathbf{Y}_{t}|\mathbf{J}^{\PP}_{t})\pi_{t-1}(\mathbf
{J}^{\PP}_{t})}{\pi_{t-1}(\mathbf{Y}_{t})},
\\
\label{below} \pi_t \bigl(\mathbf{J}^{\PP}_{t+1}
\bigr)&=&\int p \bigl(\mathbf{J}^{\PP}_{t+1}|
\mathbf{J}^{\PP}_{t} \bigr)\pi_{t} \bigl(
\mathbf{J}^{\PP}_{t} \bigr)\,d \mathbf{J}^{\PP}_{t},
\end{eqnarray}
where $f(\mathbf{Y}_{t}|\mathbf{J}^{\PP}_{t})=\prod
_{k=1}^{L}f(Y_{k,t}|\mathbf{J}^{\PP}_{t})$ and $\mathbf{Y}_{t}$ is
defined in Section~\ref{sec2}. The denominator $\pi_{t-1}(\mathbf{Y}_{t})$ in
equation (\ref{above}) is a constant, $\int f(\mathbf{Y}_{t}|\mathbf
{J}^{\PP}_{t})\pi_{t-1}(\mathbf{J}^{\PP}_{t})\,d\mathbf{J}^{\PP}_{t}$. Equation~(\ref{above}) computes the posterior density $\pi_t(\mathbf
{J}^{\PP}_{t})$ and equation (\ref{below}) is the well-known
Chapman--Kolmogorov equation, which allows us to compute the next prior
density based on $p(\mathbf{J}^{\PP}_{t+1}|\mathbf{J}^{\PP}_{t})$ [the
initial $p(\mathbf{J}^{\PP}_{0})$ is known]. For each $t$, most of the
MCMC samples are either obtained from sampling the joint $\pi_t(\mathcal
{J}^{\PP}_{t})$ or some other distribution $g_t(\mathcal{J}^{\PP}_{t})$ and\vspace*{1pt}
applying an acceptance criterion. However, the random draws of $\pi
_t(\mathcal{J}^{\PP}_{t})$ are never used again when the system proceeds
from $\pi_{t}$ to $\pi_{t+1}$ [\citet{Carlin1992}]. In high
dimensions, the posterior samples for each state will have larger
variation between iterations and, hence, both convergence and
computation problems arise. In contrast, the SIS is able to reuse the
current samples and help create the samples for the next iteration;
that improves the computational efficiency and reduces the variation
between iterations. For nonlinear problems [e.g., nonlinearity of
equation (\ref{biot})] or non-Gaussian densities, SIS requires the use
of numerical approximation techniques where the key idea is to
represent an approximation to the target posterior distribution by a
set of samples and their associated weights.

In practice, suppose a stream $S_{t}=\{(\mathcal
{J}^{\PP}_{t})^{(j)},j=1,\ldots,m\}$ ($m$~by $t$)
is a set of random samples properly weighted by the set of weights $\{w_{t}^{(j)},j=1,\ldots,m\}$ ($m$~by~1) with respect to $\pi_{t}(\mathcal
{J}^{\PP}_{t})$ [this can be viewed as approximate posterior samples of
$\mathcal{J}^{\PP}_{t}=(\mathbf{J}^{\PP}_{1},\ldots,\mathbf{J}^{\PP}_{t})$].
Define $g_{t+1}(\mathbf{J}^{\PP}_{t+1}|(\mathcal{J}^{\PP}_{t})^{(j)})$ as a
trial function for $\mathbf{J}^{\PP}_{t+1}$; the recursive SIS procedure
produces a new stream $S_{t+1}$ by drawing a new sample $\mathbf
{J}^{\PP}_{t+1}$ and updating its associated weight. This is summarized
as follows:

\begin{algorithm}[H]
\caption{SIS}\label{algo1}
\hspace*{8.2pt}(i) Sample a new $(\mathbf{J}^{\PP}_{t+1})^{(j)}$ from the trial
distribution $g_{t+1}(\mathbf{J}^{\PP}_{t+1}|(\mathcal
{J}^{\PP}_{t})^{(j)})$ and form $(\mathcal{J}^{\PP}_{t+1})^{(j)}=((\mathcal
{J}^{\PP}_{t})^{(j)},(\mathbf{J}^{\PP}_{t+1})^{(j)})$.

\hspace*{5pt}(ii) Compute the incremental weight
$u_{t+1}^{(j)}=\frac{\pi_{t+1}((\mathcal{J}^{\PP}_{t+1})^{(j)})}{\pi
_{t}((\mathcal{J}^{\PP}_{t})^{(j)})g_{t+1}(\mathbf{J}^{\PP}_{t+1}|(\mathcal
{J}^{\PP}_{t})^{(j)})}$
and update the weight $w_{t+1}^{(j)}=u_{t+1}^{(j)}w_{t}^{(j)}$.\vspace*{1pt}

(ii*) Sample a new stream $S_{t+1}'$ from the stream $S_{t+1}$ based
on the updated weights $w_{t+1}^{(j)}$.

\hspace*{2.4pt}(iii) Assign equal weights to the samples in $S_{t+1}'$.
\end{algorithm}

It has been proved that the new samples and weights $((\mathcal
{J}^{\PP}_{t+1})^{(j)},w_{t+1}^{(j)})$ are properly weighted samples from
$\pi_{t+1}$ [\citet{Liu1998}]. As time $t$ increases, a resampling
scheme is inserted between adjacent times or one can just resample
after the last time. This step is summarized in steps (ii*)~and~(iii).
\citet{Shep1997} showed that resampling [step~(ii*)] is only
necessary when the weights are very skewed; resampling reduces $m$ and
thus reduces the computational burden. A schedule for the resampling
scheme in SIS is proposed in \citet{Gordon1993}, \citet
{Kitagawa1996}\vspace*{1pt} and \citet{Liu1996}. The choice of trial
distribution $g_{t+1}(\mathbf{J}^{\PP}_{t+1}|(\mathcal
{J}^{\PP}_{t})^{(j)})$ is crucial in SIS. Choosing $g_{t+1}(\mathbf
{J}^{\PP}_{t+1}|(\mathcal{J}^{\PP}_{t})^{(j)})=\pi_{t}(\mathbf
{J}^{\PP}_{t+1}|(\mathbf{J}^{\PP}_{t})^{(j)})$ is much easier to implement,
although it might bring greater variation [see \citet
{Berzuini1997}]. This procedure ends up getting $g_{t+1}(\mathbf
{J}^{\PP}_{t+1}|(\mathcal{J}^{\PP}_{t})^{(j)})=p(\mathbf
{J}^{\PP}_{t+1}|(\mathbf{J}^{\PP}_{t})^{(j)})$ and\vspace*{1pt} incremental weights
$f(\mathbf{Y}_{t+1}|(\mathbf{J}^{\PP}_{t+1})^{(j)})=\prod
_{k=1}^{L}f(Y_{k,t+1}|(\mathbf{J}^{\PP}_{t+1})^{(j)})$.
There exist in the literature several kinds of local Monte Carlo
methods which could be embedded into SIS to get the weights or even
approximate weights no matter what $g_{t+1}$ function we choose. This
strategy provides the opportunity to find relatively good weights that
could be used in SIS so we can limit our attention to the choice of
trial function when we apply SIS. The SIS procedure (Algorithm \ref
{algo1}) was initially used in the analysis of state-space models and
is similar to
sequential Monte Carlo (SMC) which has recently been applied as an
alternative to MCMC for standard Bayesian inference problems
[\citet{Neal2001},
\citet{Moral2006}, \citet{Fearnhead2008}].

\subsection{Regular SIS method with rejection}\label{sec3.3}
This algorithm [\citet{Liu1998}] inserts the standard rejection
method as a local Monte Carlo scheme into the SIS procedure. At step
$t$, the rejection method is constructed based on sampling the joint
distribution of ($J, \mathbf{J}^{\PP}_{t+1}$). To do this, we draw $J=j$
with probability proportional to $w_{t}^{(j)}$. Given $J=j$, sample
$(\mathbf{J}^{\PP}_{t+1})^{(j)} $ from $p(\mathbf{J}^{\PP}_{t+1}|(\mathbf
{J}^{\PP}_{t})^{(j)})$. Next, compute the constant $c_{t+1}=\sup_{j}\prod
_{k=1}^{L}f(Y_{k,t+1}|\break (\mathbf{J}^{\PP}_{t+1})^{(j)})$. Then, accept
$(j,(\mathbf{J}^{\PP}_{t+1})^{(j)})$ with probability $\prod
_{k=1}^{L}f(Y_{k,t+1}|\break (\mathbf{J}^{\PP}_{t+1})^{(j)})/c_{t+1}$. Based on
the local samples from the rejection method, the estimates of the
associated weights of the samples for each state are computed by the
following procedure:
\begin{longlist}[(ii)]
\item[(i)] Estimate the weight $w_{t+1}^{(j)}$ by $\hat{f}_{j}={}$frequency
of $\{J=j\}$ in the sample.

\item[(ii)] Update the sample $(\mathcal{J}^{\PP}_{t+1})^{(j)}=((\mathbf
{J}^{\PP}_{t})^{(j)},(\mathbf{J}^{\PP}_{t+1})^{*})$ if $\hat{f}_{j}\neq0$,
where $(\mathbf{J}^{\PP}_{t+1})^{*}$ is any value of $\mathbf
{J}^{\PP}_{t+1}$ if the associated $\hat{f}_{j}\neq0$, or take a random
draw from those with $\hat{f}_{j}\neq0$ if the associated $\hat
{f}_{j}=0$.
\end{longlist}

Resample $m{'}$ out of $m$ rows from $\mathcal{J}^{\PP}_{t+1}$ without
replacement based on the weights $\{w_{t+1}^{(j)},j=1,\ldots,m\}$. In
order to improve the efficiency of SIS, the resampling scheme is used
when the SIS arrives at the last time step rather than resampling after
every step. 

\subsection{Improved SIS method with resampling}\label{sec3.4}
The disadvantage of the regular SIS with rejection method is that it
requires computing the constant $c_{t+1}$ within the embedded rejection
method and re-estimation of the weights for the SIS procedure from the
samples $\{J_{(l)},(\mathbf{J}^{\PP}_{t+1})^{(l)}\}_{l=1}^{m'}$. Both of
these computations could be quite inefficient in the state space model
with high dimension. However, an improvement could be made when the
local importance resampling takes place so that the samples are not
collected by the accept/reject ratio, but instead by assigning a weight
to each sample. Specifically, calculating the constant $c_{t+1}$ or
estimating the weights by counting $\hat{f}_{j}$ is no longer
necessary; instead we simply assign to the samples $(\mathbf
{J}^{\PP}_{t+1})^{(j)}$ the weights $w_{t+1}^{(j)}=\prod
_{k=1}^{L}f(Y_{k,t+1}|(\mathbf{J}^{\PP}_{t+1})^{(j)})$. It has been
proved [\citet{Liu1998}] that the samples from the local
importance resampling method would automatically achieve the resampling
effect. 
Thus, we could just keep those weights from any of the local Monte
Carlo methods and iterate the SIS.\looseness=-1 

\section{Simulation study}\label{sec4}
\subsection{MEG data generation}\label{sec4.1}
In a typical MEG experiment, time is measured in milliseconds (the
sampling rate is 1 kHz). However, for better understanding, from now
on, we will use timesteps rather than milliseconds. We ran two
simulated cases to verify that the methods work. First, we present some
preliminary results for the single dipole case with a few parameters
and low dimension in time. Second, an extension to the single dipole
case with six parameters and high dimension in time is given. We used
40 radially oriented magnetometers in one case, and 100 radially
oriented magnetometers in the other. The dipole was restricted to move
inside the brain. In order to focus on the source parameters, we fixed
several parameters (source noise parameters, measurement noise
parameters, etc.) in the model.

%
\begin{table}
\tabcolsep=0pt
\tablewidth=260pt
\caption{Dipole simulation: the location parameters of the dipole are
expressed in terms of Cartesian coordinates [$x$~(cm), $y$~(cm), $z$~(cm)],
$m_{1}$ and $m_{2}$ are the dipole moment
parameters. $s$~(mA) is the strength parameter of a dipole. Only
the $z$ component of the dipole is allowed to vary. The other five
components are held fixed by setting the diagonal components of the
covariance matrix to zero}\label{dipole_simu1}
\begin{tabular*}{\tablewidth}{@{\extracolsep{\fill}}@{}lc@{}} 
\hline
$\mathbf{m}_{\mathrm{int}}=(x,y,z,m_{1},m_{2},s)$ & $(1,1,5,3,3,3)$ \\
$\mathbf{m}_{\mathrm{com}}=(x,y,z,m_{1},m_{2},s)$ & $(0,0,0,0,0,0)$ \\
$\bolds{\rho}=\operatorname{diag}[\rho_{1},\rho_{2},\ldots,\rho_{6}]$ & $\operatorname{diag}[1,1,0.9,1,1,1]$ \\[2pt]
$\bolds{\Sigma}_{1}=\operatorname{diag}[\sigma_{1}^{2},\sigma_{1}^{2},\ldots,\sigma_{1}^{2}]$ & $\operatorname{diag}[0.0625,0.0625,\ldots,0.0625]$ \\[2pt]
$\bolds{\Sigma}_{2}=\operatorname{diag}[\sigma_{11}^{2},\sigma_{22}^{2},\ldots,\sigma_{66}^{2}]$ & $\operatorname{diag}[0,0,0.0225,0,0,0]$ \\
Number of timesteps & 15 \\
\hline
\end{tabular*}
\end{table}

%
\begin{figure}

\includegraphics{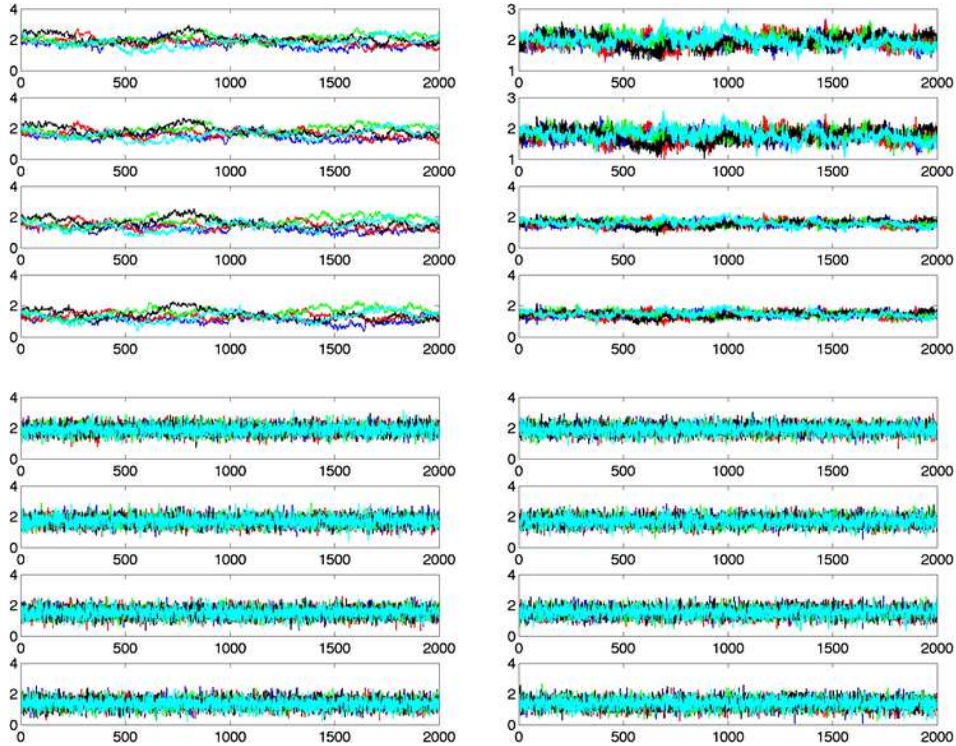}

\caption{A simple test case where only one source parameter $z$ is
allowed to vary.
Top left: trace plots of location parameter $z$ at four selected timesteps (9th, 10th, 11th and 12th) by the random-walk
MCMC within Gibbs. Similar plots are also shown for the hybrid MCMC
within Gibbs (top right), regular SIS method with rejection (bottom
left) and improved SIS method with resampling (bottom right).}\vspace*{-3pt}\label{compare}
\end{figure}

%
\begin{figure}[b]

\includegraphics{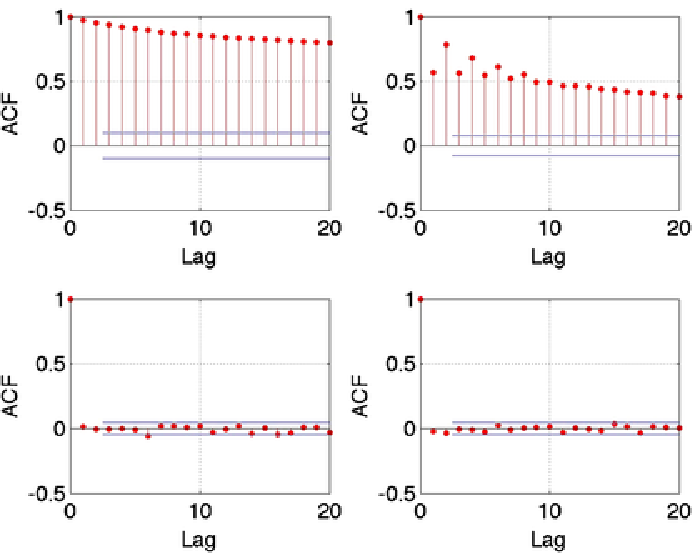}

\caption{Autocorrelation plots for the four methods.
Top left: random-walk MCMC within Gibbs;
top right: hybrid MCMC within Gibbs;
bottom left: SIS method with rejection;
bottom right: SIS with resampling.}\label{acf}
\end{figure}

\subsubsection*{Simulated case~1}
Before running our algorithms for a long time, we tested a simplified
case where the simulation was run for only $15$ timesteps with only one
of the six parameters allowed to vary. In this very simple example, the
dipole only moves in the $z$ dimension in the brain and both the
strength and moments of the dipole remain constant. The parameters of
the simulated dipole are summarized in Table~\ref{dipole_simu1}. The
regular SIS method with rejection and the improved SIS method with
resampling were tested. The random-walk MCMC within Gibbs and the
hybrid MCMC within Gibbs were also run for comparison. We randomly
generated 25 data sets and tested them under each scenario. Figure~\ref{compare} shows the trace plots (only 5 overlay plots are shown) for 4
selected timesteps from all the methods. We observe that both the
random-walk MCMC within Gibbs and hybrid MCMC within Gibbs do not
provide a stable estimate for each timepoint and their samples are
highly correlated. Both of our methods produce much nicer samples which
oscillate around the true values. To have a quantitative comparison, we
conducted a detailed convergence diagnosis for each approach: the
sample autocorrelation function of the chain at a selected timestep was
computed for each approach (see Figure~\ref{acf}); the effective sample
size of the averaged chain from each approach was calculated;
Gelman--Rubin's method was used for evaluating convergence; similar
methods such as Geweke, Heidelberger--Welch and Raftery--Lewis were also
employed for diagnosis (see Table~\ref{mcmc_diag}). Both Figures~\ref{compare},~\ref{acf} and Table~\ref{mcmc_diag} show strong evidence
that our approaches outperform the MCMC methods.

%
%
\begin{table}[t]
\tabcolsep=0pt
\caption{Convergence diagnosis by effective sample size (ESS),
Gelman--Rubin (GR) [\citet{gelmanrubin1992}]: value near 1 suggests
convergence, Geweke (GE) [\citet{Geweke92evaluatingthe}]: $z$-score
for stationary test, Heidelberger--Welch (HW) [\citet
{Heidelberger1983}]: $p$~value for stationary test, Raftery--Lewis
(RL) [\citet{rafterylewis1992}]: large value suggests strong
autocorrelation}\label{mcmc_diag}
\begin{tabular*}{\tablewidth}{@{\extracolsep{\fill}}@{}ld{4.1}d{1.3}d{3.3}d{1.3}d{2.1}@{}} 
\hline
\textbf{Method} & \multicolumn{1}{c}{\textbf{ESS}} & \multicolumn{1}{c}{\textbf{GR}}
                & \multicolumn{1}{c}{\textbf{GE}}  & \multicolumn{1}{c}{\textbf{HW}}
                & \multicolumn{1}{c@{}}{\textbf{RL}}
\\
\hline
Random-walk MCMC within Gibbs & 26.7 & 1.29 & -14.968 & 0.05 & 18.2 \\
Hybrid MCMC within Gibbs & 553.5 & 1.06 & -4.985 & 0.116 & 1.6 \\
Regular SIS with rejection & 1946.9 & 1.013 & -0.64 & 0.25 & 1.0 \\
Improved SIS with resampling & 2000.0 & 1.005 & -0.28 & 0.81 & 1.0
\\
\hline
\end{tabular*}
\end{table}

%
\begin{table}[t]
\tabcolsep=0pt
\tablewidth=275pt
\caption{Dipole simulation: the location parameters of the dipole are
expressed in terms of Cartesian coordinates [$x$~(cm), $y$~(cm), $z$~(cm)],
$m_{1}$ and $m_{2}$ are the dipole moment
parameters. $s$~(mA) is the strength parameter of a dipole. The
diagonal elements of $\bolds{\Sigma}_{1}$ and $\bolds{\Sigma}_{2}$
are 0.0625 fT$^{2}$ and 0.01~cm$^{2}$, respectively}\label{dipole_simu2}
\begin{tabular*}{\tablewidth}{@{\extracolsep{\fill}}lcc@{}} 
\hline
Initial timepoint & \\
\quad $\mathbf{m}_{\mathrm{int}}=(x,y,z,m_{1},m_{2},s)$ & $(6,7,8,3,5,5)$ \\
\quad $\mathbf{m}_{\mathrm{com}}=(x,y,z,m_{1},m_{2},s)$ & $(0,0,0,0,0,0)$ \\
\quad $\bolds{\rho}=\operatorname{diag}[\rho_{1},\rho_{2},\ldots,\rho_{6}]$ & $\operatorname{diag}[0.65,0.7,0.75,0.8,0.85,0.9]$ \\[2pt]
Random-walk move& \\
\quad $(x,y,z,m_{1},m_{2},s)$ & \mbox{Based on previous value} \\
\quad Number of timesteps & 10
\\[2pt]
Autoregressive move & \\
\quad $(x,y,z,m_{1},m_{2},s)$ & \mbox{Based on previous value} \\
\quad $\mathbf{m}_{\mathrm{com}}=(x,y,z,m_{1},m_{2},s)$ & $(0,0,0,0,0,0)$ \\
\quad $\bolds{\rho}=\operatorname{diag}[\rho_{1},\rho_{2},\ldots,\rho_{6}]$ & $\operatorname{diag}[0.65,0.7,0.75,0.8,0.85,0.9]$\\[2pt]
Random-walk move& \\
\quad $(x,y,z,m_{1},m_{2},s)$ & \mbox{Based on previous value} \\
\quad Number of timesteps & 10 \\[2pt]
$\cdots$& $\cdots$\\
Repeat until 100th timepoint & \\
\hline
\end{tabular*}
\end{table}

\begin{figure}

\includegraphics{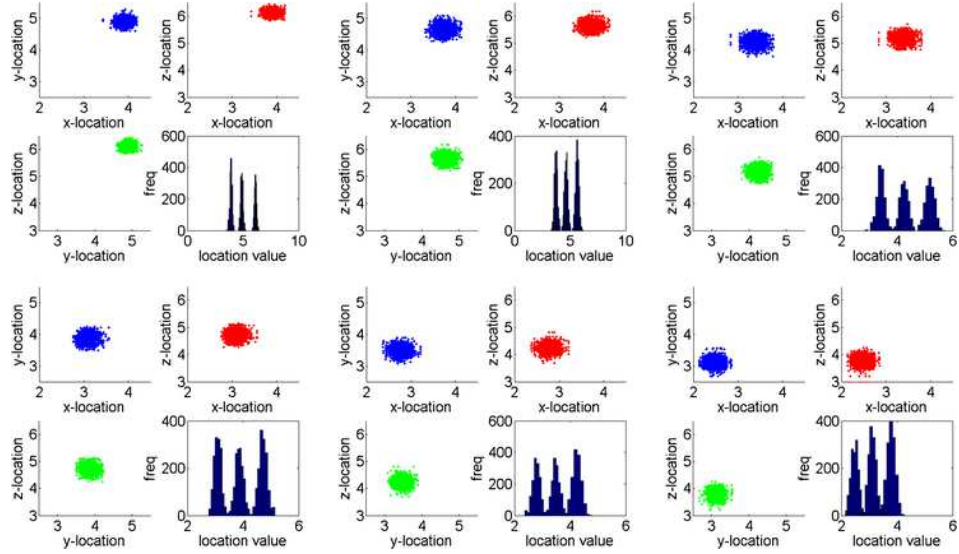}

\caption{Source location parameters at six timesteps (a total of six
$2\times2$ subplots).
Top left: in this $2 \times2$ subplot, there are
three pairwise plots of the source location parameters ($x$~and~$y$,
$x$~and~$z$, $y$~and~$z$) at 1st timestep and one side by side
histogram plot for the source location parameters ($x$, $y$ and~$z$) at
1st timestep. The rest of the five subplots give the same information
for different timesteps: 20th timestep (top middle), 40th timestep (top
right), 60th timestep (bottom left), 80th timestep (bottom middle) and
100th timestep (bottom right).}\vspace*{-3pt}\label{six_para_1}
\end{figure}

\subsubsection*{Simulated case 2}
\begin{figure}

\includegraphics{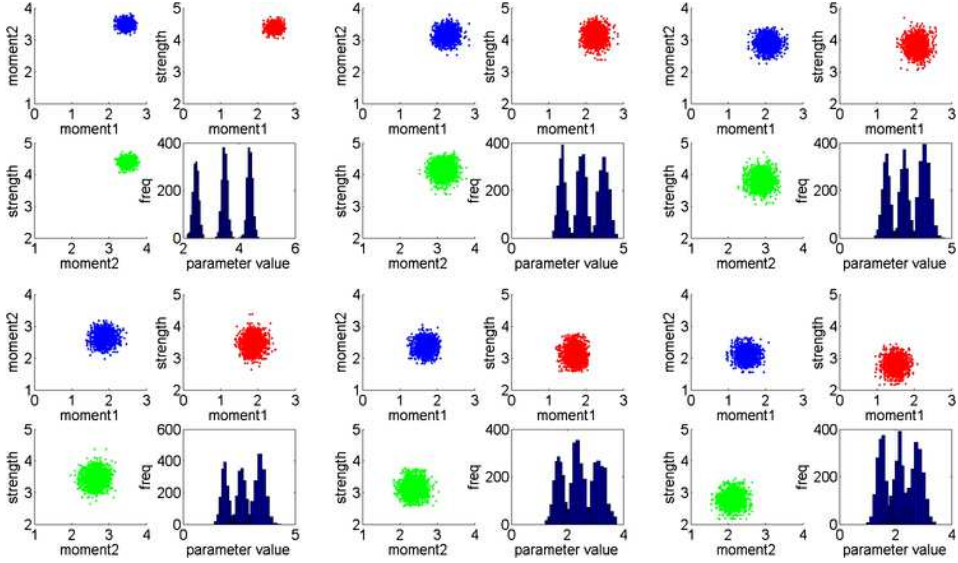}

\caption{Source moment and strength parameters at six timesteps (a
total of six $2\times2$ subplots).
Top left: in this $2 \times2$
subplot, there are three pairwise plots of the source moment and
strength parameters ($m_{1}$ and $m_{2}$, $m_{1}$ and $s$ and $m_{2}$
and $s$) at 1st timestep and one side by side histogram plot for the
moment and strength parameters ($m_{1}$, $m_{2}$ and $s$) at 1st
timestep. The rest of the five subplots give the same information for
different timesteps: 20th timestep (top middle), 40th timestep (top
right), 60th timestep (bottom left), 80th timestep (bottom middle) and
100th timestep (bottom right).}\label{six_para_2}
\end{figure}

In addition to case 1, a case of multiple-source parameters (three
location parameters and three moment and strength parameters) was
performed. In this simulation, the source was modeled as a moving
dipole following a multivariate autoregressive time series. The dipole
moves in the three coordinate directions $x$, $y$ and $z$, and both
strength and moments of the dipole change as well. The total length of
simulation is 100 timesteps (we will run 2000 timesteps for data in
Section~\ref{sec4.3}). To control the movement of the simulated dipole (to not
move outside of the brain when the number of timepoints are large), we
restricted the range of each parameter for the dipole. In order to do
this, we set boundary values for each parameter (i.e., maximum and
minimum). The autoregressive model for $\mathbf{J}^{\PP}_{t}$ in Section~\ref{sec2}
occurred only at certain timepoints when specified in advanced. In
other words, the dipole had two types of moves: one is a move based on
the autoregressive model, and the other is a random-walk move. The
dipole moved according to the autoregressive model at certain specified
timesteps whereas the random walk was applied to the dipole at the rest
of the timepoints. We had similar restrictions on the other parameters
of the dipole. The parameters setup is given in Table~\ref{dipole_simu2}. The plots (histogram) for each dipole location
parameter and pairwise plots for the location parameters are shown in
Figure~\ref{six_para_1}. These side by side histograms show the
distribution of each location parameter at six selected timepoints.
Similar plots for the other three moment and strength parameters are
also shown in Figure~\ref{six_para_2}. We can see that the
distributions (non-Gaussian) of each parameter of the source are
varying at each timestep as we expected.
%


%

\subsection{Parallel virtual machine (PVM) for high dimension in time}\label{sec4.2}
In practice, the MEG data set we have from an experiment is very large
(e.g., hundreds of thousands of timesteps). The same algorithms
(Sections~\ref{sec3.3} and~\ref{sec3.4}) need to be run for a much longer time. To
be exact, if we run for 5000 timesteps with 1500 replications (sample
paths) for each $\mathbf{J}^{\PP}_{t}$, we are supposed to get a stream
of $S_{5000}=\{(\mathcal{J}^{\PP}_{5000})^{(j)}, j=1,\ldots, 1500\}$
($S_{t}$ is defined in Section~\ref{sec3.2}). Because of the sequential
character of our algorithms, sample paths [$(\mathbf{J}^{\PP}_{t})^{(j)},
j=1,\ldots,m$] for each time are computed in a sequential fashion and
the weights updated at each time. Therefore, it is very inefficient to
get the sample paths for a longer time.


Note that we always need the sample path from the previous time
($\mathbf{J}^{\PP}_{t-1}$) when we work on the current time ($\mathbf
{J}^{\PP}_{t}$) and they are not independent, therefore, it is not
possible to improve the speed in the direction of time [e.g., $(\mathbf
{J}^{\PP}_{t})^{(j)}$ sequentially depends on $(\mathbf
{J}^{\PP}_{t-1})^{(j)}$]. However, the sample paths are independent
within each timestep; this is to say, at time $t$, $(\mathbf
{J}^{\PP}_{t})^{(j)}$ is independent $(\mathbf{J}^{\PP}_{t})^{(j')}$, so
they can be computed in a separate fashion. In other words, it is
always possible for us to compute several sample paths (several chunks)
for the same timestep (at time~$t$) simultaneously. This simultaneous
computation for sample paths up to the final timestep (5000) could be
achieved by parallel computing where each parallel thread would contain
a sequential calculation for all the time $t$ ($1 \leq t \leq5000$)
with fewer samples, so that our sequential problem can be solved in parallel.
The Parallel Virtual Machine (PVM) [\citet{PVM1}], a parallel
computing paradigm, is used to speed up the computation. It is designed
to allow a network of heterogeneous machines to be used as a single
distributed parallel processor, so that a large scale computing problem
can be solved more cost effectively. The PVM structure we use is a
Master--Worker model where there are several worker programs performing
tasks in parallel and a master program collecting the outcomes from
each worker. Each task is to separately compute a subset of the sample
paths for all timesteps. The resampling scheme is included in the
worker program and there is no parallelism in time. To be exact, if
there are three worker programs in the Master--Worker model to generate
a steam $S_{T}=\{(\mathcal{J}^{\PP}_{T})^{(j)}, j=1,\ldots,m\}$, the way
of running PVM is as follows:

\begin{algorithm}[H]
\hspace*{3pt}(i) Initialize each worker program and let each worker run for a substream
$S'_{T}=\{(\mathcal{J}^{\PP}_{T})^{(j)}, j=1,\ldots,\frac{m}{3}\}$.

(ii) Stack each $S'_{T}$ and get a complete $S_{T}$.
\caption{PVM schedule}
\end{algorithm}

The size of $S'_{T}$ can be adjusted according to the size of
$S_{T}$ and the number of worker programs that are in use. The speed is
mainly influenced by hardware and software components of network and
I/O systems. It also depends on the number of worker programs, for
example, adding too many parallel workers does not enhance the speed
when most of the time is spent on master--worker communication. In
practice, deciding on the number of workers requires experience and it
varies for different machines. Because the magnetic fields generated by
independent dipoles add up, there is no additional complexity (other
than increased computation) brought by multiple dipoles.

Since our PVM program involves randomness and a resampling scheme,
several issues still need to be resolved. First, if our algorithms were
implemented in a single program without parallelism, all samples
generated before resampling from this program should be simply related
to the random number generator. However, when there are several
workers, each of them doing the same thing as a single program but in
parallel, the unique randomness within each worker will eventually come
up with different but similar samples before resampling. To be exact,
in order to have the two programs generate the same results, in the PVM
structure we need to explicitly and precisely choose different workers
according to a predefined random sequence. This random sequence can be
obtained from a single program without parallelism. Unfortunately, this
needs a lot of work in programming and would surely slow down the
computation. Second, in a single program without parallelism, we would
only have one resampling procedure. The samples would be generated from
the resampling procedure. However, there would be one resampling
procedure within each of our worker programs in PVM. The samples would
be generated from each of these workers and should eventually be pooled
together. In principle, the weights from each worker should be pooled
first and then we would perform the resampling procedure. The reason is
that each worker might generate different weights so that the
normalizing constants might be different. If the resampling happens
only one time (at the end of all timesteps), a reasonable way to solve
this problem is that we can do the resampling scheme in the master
program after normalizing all the weights when pooled. If there were
several resampling schemes before the end, we could still return to the
master program when necessary. Again, this needs extensive programming
and, again, it would surely slow down the computations. In our current
program, sums of weights within each worker were almost the same
(normalizing constants were almost the same), so we retained the
resampling procedure in each worker program. Because the random number
generation will not produce the same numbers in a parallel program as
in a sequential program without extensive interprocess communication
and because resampling within each parallel worker program will produce
different results than would resampling in the master, we do not expect
the identical samples in the parallel version of our sequential
program. We do expect the distribution of the samples from the parallel
program to be indistinguishable from the distribution of the samples
from the sequential version.


\subsection{Numerical results for running PVM for MEG model}\label{sec4.3}
The PVM was first run on a single Linux workstation (Intel Pentium 4
CPU 3.80~GHz, Memory 2 GB) for different configurations. The data size
was $2000$ MEG timesteps with $1500$ sample paths for each timestep. We
split the computation into a number of tasks: $1$ (without PVM), $3$,
$5$, $10$ and $15$ workers, respectively, and run for $100$, $500$,
$1000$, $1500$ and $2000$ timesteps. The user CPU time (total number of
CPU-seconds for master and worker programs) is used to measure the time
spent by each PVM run. The real time elapsed (minutes) is also shown in
parentheses beside the user CPU time. The result is shown in Table~\ref{single_PVM}.
%
%
\begin{table}[b]
\tabcolsep=0pt
\caption{PVM application on a single workstation. Five different PVM
configurations were run. The number of workers in PVM is denoted
``number of tasks.'' The number of sample paths within each worker is
denoted ``load per task.'' Each PVM run eventually generates 1500
sample paths}\label{single_PVM}
\begin{tabular*}{\tablewidth}{@{\extracolsep{\fill}}ld{4.0}ccccc@{}} 
\hline
\textbf{Number} & \multicolumn{1}{c}{\textbf{Load}} & \textbf{Time 1} & \textbf{Time 2} & \textbf{Time 3} & \textbf{Time 4} & \textbf{Time 5} \\
\textbf{of tasks} & \multicolumn{1}{c}{\textbf{per task}} & \textbf{(100)} & \textbf{(500)} & \textbf{(1000)} & \textbf{(1500)} & \textbf{(2000)}\\
\hline
\phantom{0}1 & 1500 & 0.008 (1.00) & 0.032 (5.12) & 0.064 (10.35) & 0.120 (16.47) & 0.136 (22.08) \\ 
\phantom{0}3 & 500 & 0.008 (0.24) & 0.032 (2.05) & 0.060 (4.12)\phantom{0} & 0.096 (6.23)\phantom{0} & 0.148 (8.40)\phantom{0} \\
\phantom{0}5 & 300 & 0.008 (0.17) & 0.036 (1.27) & 0.064 (3.17)\phantom{0} & 0.104 (4.25)\phantom{0} & 0.148 (6.47)\phantom{0}\\
10 & 150 & 0.008 (0.11) & 0.040 (0.59) & 0.072 (1.59)\phantom{0} & 0.096 (3.00)\phantom{0} & 0.136 (4.51)\phantom{0}\\
15 & 100 & 0.008 (0.10) & 0.036 (0.50) & 0.064 (1.43)\phantom{0} & 0.124 (2.33)\phantom{0} & 0.164 (3.24)\phantom{0}\\  
\hline
\end{tabular*}
\end{table}

\begin{figure}[t]
\begin{tabular}{@{}c@{\qquad}c@{}}

\includegraphics{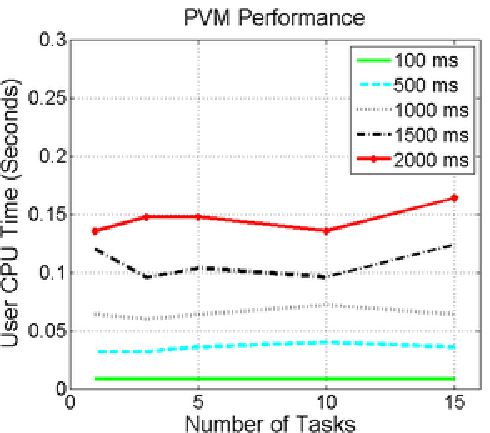}
 & \includegraphics{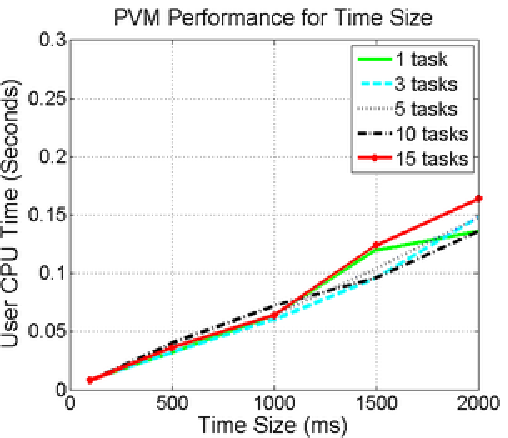}\\
\footnotesize{(a) CPU time with tasks} & \footnotesize{(b) CPU time with timesteps}
\end{tabular}
\caption{PVM Performance: user CPU time (seconds) for number of tasks
and different time run.
\textup{(a)}~Each line (with a specific timestep) is a plot of user CPU time for different number of tasks.
\textup{(b)}~Each line (with a specific number of tasks) is a plot of user CPU time for different timesteps.}\label{PVM_perform1}
\end{figure}

We can see that the user CPU time increases roughly linearly in the
number of timesteps from $0.008$ second to $0.146$ second on average.
The linear relationship of user CPU time on experiment time is almost
the same for each of these PVM configurations as we expected. This can
be clearly observed from Figure~\ref{PVM_perform1}: in Figure~\ref{PVM_perform1}(a), these lines (user CPU time/Task) are nearly equally
distant and stay roughly constant for different tasks within
the\vadjust{\goodbreak}
samesteps time run; in Figure~\ref{PVM_perform1}(b), the slope of each
line (user CPU time/Timesteps) is almost the same. Note that there is a
significant difference in real time elapsed for different PVM
configurations. This should not be considered a contradiction with user
CPU time because real time elapsed is mostly affected by other programs
and it includes time spent in memory, I/O and other resources.

%


%
%
\begin{table}[b]
\tabcolsep=0pt
\caption{PVM application on multiple workstations. This table shows the
user CPU time (seconds) for each PVM run and real time elapsed
(minutes) in parentheses using one, two, three and four machines. The
length of each PVM run was 1500 timesteps}\label{multiple_PVM}
\begin{tabular*}{\tablewidth}{@{\extracolsep{\fill}}@{}lccccc@{}} 
\hline
\textbf{Number} & \textbf{Load} & \textbf{Time 1} & \textbf{Time 2} & \textbf{Time 3} & \textbf{Time 4}\\
\textbf{of tasks} & \textbf{per task} & \textbf{(1500)} & \textbf{(1500)} & \textbf{(1500)} & \textbf{(1500)} \\
\hline
\phantom{0}3 & 500 & 0.084 (7.56) & 0.128 (5.46) & 0.108 (3.39) & 0.096 (2.34)  \\
\phantom{0}5 & 300 & 0.100 (4.55) & 0.084 (3.10) & 0.124 (2.23) & 0.108 (1.29)  \\
10 & 150 & 0.100 (3.19) & 0.096 (1.51) & 0.104 (1.39) & 0.116 (1.03)  \\
15 & 100 & 0.124 (2.34) & 0.112 (1.31) & 0.104 (1.00) & 0.112 (0.44)  \\
\hline
\end{tabular*}
\end{table}

\begin{figure}
\begin{tabular}{@{}c@{\qquad}c@{}}

\includegraphics{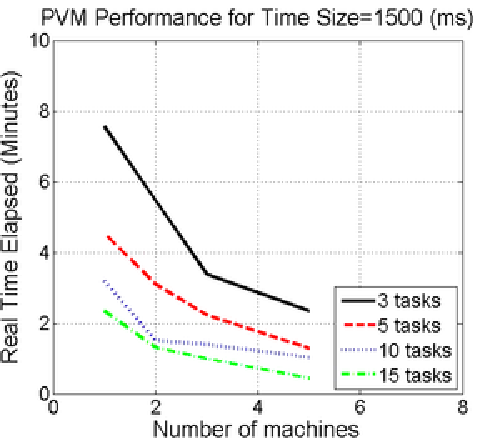}
 & \includegraphics{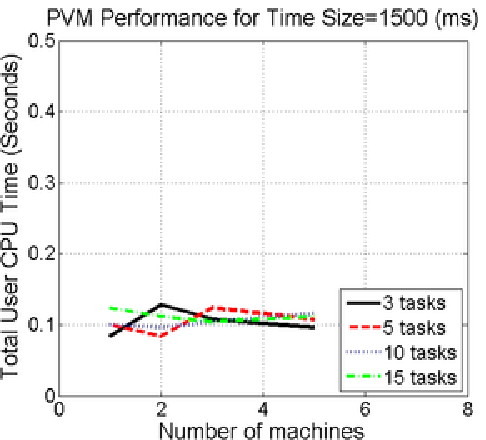}\\
\footnotesize{(a) Real time for PVM run} & \footnotesize{(b) Total user CPU time for PVM run}
\end{tabular}
\caption{PVM Performance: real time elapsed (minutes) and user CPU time
(seconds) graph for number of machines for 1500 timesteps PVM run.
\textup{(a)}~Each line (with a specific number of tasks) is a plot of real time elapsed for a different number of machines.
\textup{(b)}~Each line (with a specific number of tasks) is a plot of total user CPU time of master and worker programs for different number of machines.}\label{PVM_perform2}
\end{figure}

The performance can still be improved when extra machines are included.
Table~\ref{multiple_PVM} lists the PVM performance of 1--4 machines with
$1500$ timesteps. First, since user CPU time is the sum of the CPU time
for master and worker programs, it is expected that the user CPU time
for each of these PVM runs is roughly $0.120$ second. Second, the real
elapsed time of each PVM run is cut to 50\%--70\% if one machine is
added. It then goes down to 40\%--50\% when three computers are
employed. The real time elapsed decreases to 10\%--30\% when four
computers are added. These performances are based on our public cluster
with heterogeneous CPU speed and cache size. The theoretical reduction
in execution time of PVM is not necessarily achieved. Finally, to get
better time execution by PVM, we suggest to adjust the number of CPUs
and the number of tasks, and to use relatively similar machines. To
summarize, Figure~\ref{PVM_perform2} is a graphic illustration of both
real time elapsed and user CPU time for our PVM run.

%
%


\section{A real data application}\label{sec5}

\begin{figure}[t]

\includegraphics{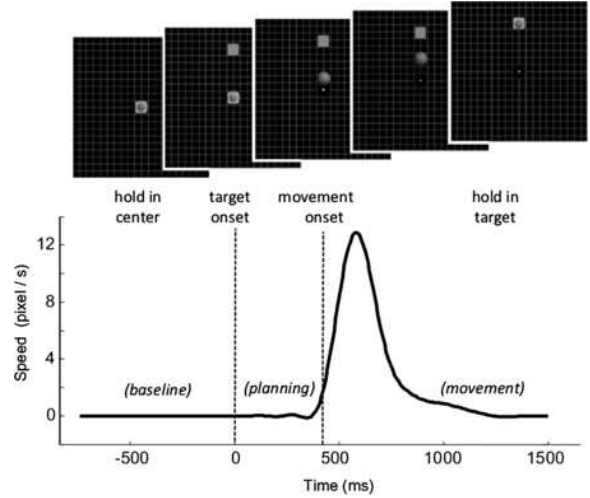}

\caption{The subject controls the 2-D cursor position using wrist
movements. The cursor needs to go to the center and stay there for a
hold period until the peripheral target appears. Then the cursor moves
from the center out to the target and stays there for another hold
period to complete the trial successfully. The target changes color
when hit by the cursor and disappears when the holding period has
finished. The bottom trace shows the speed profile of the cursor from a
representative trial, and the dotted lines delimit the
pre-movement/planning period. Figure and explanation were obtained from \citet{Wei2010}.}\label{experiment}
\end{figure}

Data was collected by a 306-channel system (Elekta-Neuromag) at the
Center for Advanced Brain Magnetic Source Imaging (CABMSI) at UPMC
Presbyterian Hospital in Pittsburgh in an experiment related to
Brain-controlled interfaces (BCI). A BCI expresses motor commands via
neural signals directly from the brain. The experiment involves two
parts (see Figure~\ref{experiment}): in the first part the subjects
were asked to imagine performing the ``center-out'' task using the
wrist (imagined movement task) and in the second part the subjects
controlled a 2-D cursor using the wrist to perform the center-out task
following a visual target (overt movement task). 

%
%
%
\begin{figure}[b]

\includegraphics{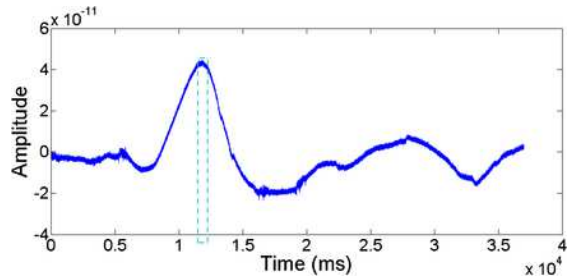}

\caption{The MEG signal of a typical trial at a magnetometer. The
horizontal axis is time (ms) and the vertical axis is the magnitude of the signal (fT).}\label{dashbox}
\end{figure}

\begin{figure}[t]

\includegraphics{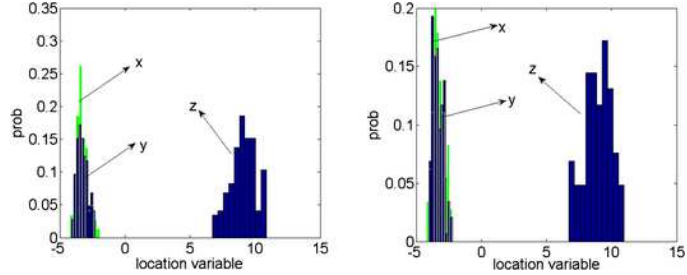}

\caption{Empirical density for source location parameter ($x$,
$y$, $z$). Left: density plot for $1$st millisecond;
right: density plot for $101$st millisecond.}\label{emdensity}
\end{figure}

\begin{figure}[b]

\includegraphics{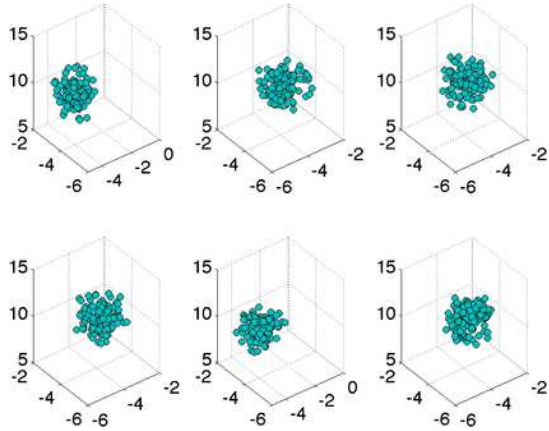}

\caption{Time variation of the dipole distribution. Upper row: dipole
distribution at $1$st, $21$st and $41$st millisecond (left to right);
lower row: dipole distribution at $61$st, $81$st and $101$st millisecond.}\label{3dplot}
\end{figure}

The data consists of one trial recording 37,000 milliseconds at 102 MEG
sensors (magnetometers). We used this data for testing our model along
with our PVM scheme. Instead of analyzing the whole trial of data, we
only analyzed about 400 milliseconds (dashed box in Figure~\ref{dashbox}) after movement onset (12,000~milliseconds--12,400 milliseconds in the original data) from all the channels. To simplify
our calculation for the real data, we were only estimating the location
of the source $(x,y,z)$. The moment and strength parameters
$(m_{1},m_{2},s)$ were not of our interest (not varying too much by
assumption). The choice of prior for real data is an open question; we
used almost the same prior as we did in Section~\ref{sec4} for simplification.
We set the mean $\mathbf{m}_{\mathrm{ini}}$ of the initial state $\mathbf
{J}^{\PP}_{0}$ as $(-4,-4,11)$ 
motivated by the minimum norm estimate [\citet{Hamalainen1994}],
which is $(-2,-2,10)$. We further assumed a unit moment and strength
for the dipole in the data. The starting values for the initial state
$(x, y, z, m_1, m_2, s)$ were set to
$(-4.06,-3.77,13.13,1.11,0.98,1.12)$. The empirical density plots of
the dipole location parameter ($x$, $y$, $z$) at
two selected timesteps are shown in Figure~\ref{emdensity}. Using the
density plots of the location parameters, we were able to find the
dipole distribution at different timesteps. Figure~\ref{3dplot} shows
several snapshots of dipole distribution at six timesteps, that is, the
data cloud in each plot tells where the dipole might be located at a
given timestep. A full movie of the dipole distribution for 100
milliseconds can be found at
\url{http://smat.epfl.ch/\textasciitilde zyao/3dplot_animation.gif}. Different initial
values might have different performance due to the complexity of the
problem and the real data, thus, a more realistic prior needs to be
investigated in our future work. We ran PVM for 1500 milliseconds
(12,000 milliseconds--13,500 milliseconds in the original data) with the
same PVM configuration as our simulation; the time spent was very close
to that from our previous simulation results.

%
%

A typical MEG analysis would report the estimated movement of the
dipole at each time step. We can do the same. Additionally, because we
have samples from a probability distribution we can provide estimates
of the variation of the estimated movement and other source parameters.
Other methods cannot provide appropriate estimates of variation. In
clinical applications, estimates of variation may provide neurosurgeons
a much better basis for their decisions. We still need to overcome the
computational burden to make our method feasible for a complete application.

\section{Conclusion and discussion}\label{sec6}
We have introduced a general state-space formulation for the time
dependency of the parameters of a current dipole which generates the
MEG signals. In this paper we have only considered the simplest sort of
model, a first order autoregression for the time dependency. However,
the framework allows for much more complex models. And, because three
of the parameters are spatial coordinates, the model automatically
incorporates space--time dependencies. The time dependency in the model
has greatly expanded the parameter space. That fact together with the
nonlinearity of the model means that more typical MCMC methods converge
extremely slowly. The benefit of sequential methods is that we do not
attempt to estimate the entire target distribution at once but rather
attempt to estimate samples for each time point sequentially. Because
of the expanded parameter space, the need for parallel computational
methods is obvious. Our initial attempt utilized PVM and provided the
expected reduction in running time.

The results so far are mainly based on a one-source model where we
assumed there was only one dipole in the data. The extension from one
source to multiple sources is natural and only the computational
complexity increases. Our algorithms will still work in this
multiple-source model case. However, to determine the number of sources
in the MEG data is still an open question. In general, there are three
ways of finding the number of sources for the advanced model. The first
one, which is relatively easy, is to use a predefined number of sources
for the data. The second one is to estimate the number of sources from
the data in advance [\citet{Waldorp2005},
\citet{Bai2006}, \citet{Yao2012}]. The third one is to model
the number of the sources using a prior distribution [\citet
{Bertrand2001}].

We fixed several parameters when we compared our algorithms with other
MCMC methods. In fact, those parameters could be estimated along with
the source distribution. The natural way of implementing this is to
include the estimation of those parameters and the source distribution
in the iterations until all of them become stable. Furthermore, the
skewness of weights that arises in sequential importance sampling could
be a trade-off between the efficiency of the algorithm and the quality
of the source distribution. Naturally, we have observed some skewness
in the weights; we do not have enough experience to evaluate whether
this skewness should be considered excessive or unusual. Residual
sampling [\citet{Liu1998}] or stratified sampling [\citet
{Kitagawa1996}] could replace regular weight sampling and might address
excessive skewness.

To summarize, due to its nonuniqueness, finding a good estimate of the
MEG source is a challenging problem which is still open. We have
proposed a predictive model for finding a distribution for the MEG
source and we have applied our methods on both simulated data and real
data. In practice, the MEG data sets from different experimental
settings are much more complicated. Our methods can be used as a
reference with other source localization methods. Driven by the desire
of looking at the brain activity in real time, we plan to implement a
computing environment to study the brain activity under the real MEG
temporal resolution ($1/1000$~sec). The computational challenge arises
due to the extremely large dimensionality of the problem (high
resolution); there is no common computing architecture that could help.
We are exploring the use of more advanced forms of parallelism such as
CUDA (Compute Unified Device Architecture) and OPENCL to further reduce
the running time in the future.

\section*{Acknowledgments}
We thank Rob Kass and his collaborators for allowing us to use their
BCI data to test our methods. Leon Gleser gave helpful comments on the
paper. We thank the referee, the Area Editor and the Editor-in-Chief
for their helpful comments. We are especially grateful to the Associate
Editor for his/her extremely helpful comments on versions of the manuscript
and for his patience and persistence.
We would also like to thank Dr. Alberto Sorrentino and Prof. Michele Piana of the
Dipartimento di Matematica, Universita di Genova for providing several references.

\printaddresses

\end{document}